\documentclass[twocolumn,showpacs,prl,amsmath,amssymb]{revtex4}
\usepackage{graphicx}
\usepackage{dcolumn}
\usepackage{bm}

\begin{document}
\title{Kohn Anomalies and Electron-Phonon Interaction in Graphite}

\author{S. Piscanec$^1$, M. Lazzeri$^2$, Francesco Mauri$^2$,
A. C. Ferrari$^1$, and J. Robertson$^1$}
\affiliation{$^1$Cambridge University, Engineering Department,
Trumpington
Street, Cambridge CB2 1PZ, UK\\
$^2$Laboratoire de Min\'eralogie-Cristallographie de Paris, 4
Place Jussieu, 75252, Paris cedex 05, France}
\date{\today}

\begin{abstract}
We demonstrate that graphite phonon dispersions have two Kohn
anomalies at the ${\bm \Gamma}$-E$_{2g}$ and ${\bf K}$-A$'_1$
modes. The anomalies are revealed by two sharp kinks. By an exact
analytic derivation, we show that the slope of these kinks is
proportional to the square of the electron-phonon coupling (EPC).
Thus, we can directly measure the EPC from the experimental
dispersions. The ${\bm \Gamma}$-E$_{2g}$ and ${\bf K}$-A$'_1$ EPCs
are particularly large, whilst they are negligible for all the
other modes at ${\bm \Gamma}$ and ${\bf K}$.
\end{abstract}

\maketitle Carbon nanotubes (CNTs) are at the core of
nanotechnology research. They are prototype one dimensional
conductors. Metallic nanotubes are predicted to be one-dimensional
quantum wires with ballistic electron transport~\cite{todorov}.
However, high field electrical transport measurements show that
the electron-phonon scattering by optical phonons at ${\bf K}$ and
${\bm \Gamma}$ breaks down the ballistic
behavior~\cite{nanotubes}. Electron phonon coupling (EPC) is thus
the fundamental bottleneck for ballistic transport. Raman
spectroscopy is a most used characterization technique to identify
CNTs in terms of their size and electronic properties~\cite{Rao}.
The optical phonons at ${\bf K}$ and ${\bm \Gamma}$ are the
phonons responsible for the Raman D and G peaks in
carbons~\cite{Ferrari00}. The frequency and the intensity of the
Raman modes are determined by the EPC matrix
elements~\cite{Thomsen00}. The determination of the EPCs is
necessary to settle the 35 years debate on the nature of the Raman
D peak in carbons
~\cite{Tuinstra,Mapelli,Ferrari00,Maultzsch04,Pocsik,Thomsen00,Sait,Matt,Grun}.
Finally, although graphite phonon dispersions have been widely
studied, several contrasting theoretical dispersions were
proposed~\cite{Mapelli,Pavone,Maultzsch04,Matt,Grun,Sait,Dubay}.
In particular, the origin of the large overbending of the ${\bf
K}$-A$'_1$ branch is not yet understood and is associated to an
intense EPC~\cite{Ferrari00,Mapelli,Maultzsch04}. In principle,
the electronic and vibrational properties of CNTs can be described
by folding the electronic and phonon dispersions of graphite. The
precise determination of the graphite EPCs is thus the crucial
step to understand the properties of any carbon based material and
CNTs in particular. It is then surprising that, despite the vast
literature on carbon materials and CNTs, no experimental
determination or first principle calculations of the graphite EPCs
has been done so far, to the best of our knowledge.
\\\indent
Here we show that in graphite the EPC matrix elements at ${\bm
\Gamma}$ and ${\bf K}$ can be directly extracted from the phonon
dispersions. We demonstrate two remarkable Kohn anomalies in the
phonon dispersions at ${\bm \Gamma}$ and ${\bf K}$, by an exact
analytic derivation and accurate density functional theory (DFT)
calculations. We prove that the slope of the anomalies is
proportional to the EPC square.
\\\indent
A key feature of graphite is the semi-metallic character of the
electronic structure. In general, the atomic vibrations are
partially screened by electrons. In a metal this screening can
change rapidly for vibrations associated to certain ${\bf q}$
points of the Brillouin Zone (BZ), entirely determined by the
shape of the Fermi surface. The consequent anomalous behavior of
the phonon dispersion is called Kohn anomaly~\cite{Kohn59}. Kohn
anomalies may occur only for wavevectors ${\bf q}$ such that there
are two electronic states ${\bf k}_{1}$ and ${\bf k}_{2}={\bf
k}_{1}+{\bf q}$ both on the Fermi surface~\cite{Kohn59}. The
electronic bands dispersions of graphite are, essentially,
described by those of an isolated graphene sheet. In graphene, the
gap between occupied and empty electronic states is zero only at
the two equivalent BZ points ${\bf K}$ and ${\bf K}'$. Since ${\bf
K}'=2{\bf K}$, the two equivalent ${\bf K}$ points are connected
by the vector ${\bf K}$. Thus, Kohn anomalies can occur for ${\bf
q}={\bm \Gamma}$ or ${\bf q}={\bf K}$.
\\\indent
We perform calculations within the generalized gradient
approximation (GGA)~\cite{PBE}, using the density functional
perturbation theory (DFPT) scheme~\cite{DFPT}, which allows the
exact (within DFT) computation of phonon frequencies at any BZ point.
We use the plane-waves ($90~{\rm Ry}$ cut-off) and
pseudopotential~\cite{Pseudo} approach.
The electronic levels are occupied with a finite {\it fictitious}
electronic temperature $\sigma$~\cite{Methfessel}. This smears
out the discontinuities present in the Fermi distribution
for $\sigma\!=\!0$, and the exact result is recovered for
$\sigma\!\rightarrow\!0$.
The experimental lattice ($a_{\rm exp}=2.46$~\AA, $c=6.708$~\AA)
is used for graphite, while for
graphene we consider both the graphite $a_{\rm exp}$ and the
theoretical values ($a_{\rm
th}=2.479$~\AA). Graphene layers are separated by 7.4~\AA~of
vacuum.
\begin{figure}
\centerline{\includegraphics[width=85mm]{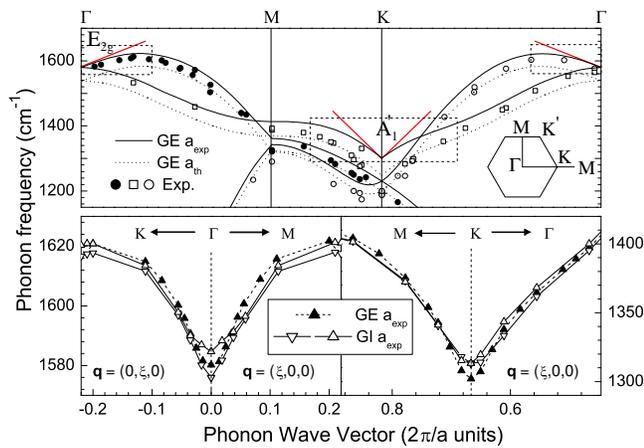}}
\caption{(Color online) Upper panel: Lines are the phonon
dispersion of graphene (GE), calculated at the experimental and
equilibrium lattice spacings (a$_{\rm exp}$ and a$_{\rm th}$).
Experimental data from Ref.~\cite{Maultzsch04}. The red straight
lines at ${\bm \Gamma}$ and ${\bf K}$ are obtained from
Eqs.~\protect\ref{eq10},\protect\ref{eq12}. The two lower panels
correspond to the dotted window in the upper panel. Here, graphite
(GI) computed frequencies are also shown. The points are
theoretical frequencies obtained by direct calculation. A single
GE band corresponds to two almost-degenerate GI-bands.
}\label{fig1}
\end{figure}
\\\indent
Fig.~\ref{fig1} compares the measured optical
branches~\cite{Maultzsch04} with our calculations at
$\sigma=0.02~{\rm Ry}$. Phonon frequencies are computed exactly
for a series of points along the high symmetry lines ${\bm
\Gamma}$-${\rm \bf K}$ and ${\bm \Gamma}$-${\rm \bf M}$ and then
interpolated with a spline. The agreement with experiments is
$\sim 2\%$, which is the expected accuracy of DFT-GGA. At ${\bm
\Gamma}$ the experimental dispersion is very well reproduced by
the calculations with $a_{\rm exp}$. At ${\rm \bf K}$ the upper
branch is better described by the calculations with $a_{\rm th}$.
\\\indent
The most striking feature of these dispersions is the
discontinuity in the frequency derivative of the highest optical
branches (HOB) at ${\bm \Gamma}$ and at ${\rm \bf K}$ (E$_{2g}$
and A$'_1$ modes). Indeed, near ${\bm \Gamma}$, $\hbar\omega_{\bf
q}=\alpha_{\bm \Gamma} q +\hbar\omega_{\bm \Gamma}+ {\cal
O}(q^2)$, $\omega_{\bf q}$ being the phonon frequency of the HOB
at the ${\bf q}$ wavevector. Similarly, near ${\bf K}$,
$\hbar\omega_{\bf  K+q'}=\alpha_{\bf K} q' +\hbar\omega_{\bf K}+
{\cal O}(q'^2)$. Such dependencies cannot be described by a finite
set of interatomic force constants, or by a set
decaying exponentially with the real-space distance. In these
cases the dynamical matrix dependence on ${\bf q}$ would be
analytic, and, because of symmetry, {\it the highest optical
branch near ${\bm \Gamma}$ and ${\rm \bf K}$ would have a flat
slope} ($\alpha_{\bm \Gamma}\!=\!\alpha_{\bf K}\!=\!0$). Thus, a
non zero $\alpha_{\bm \Gamma}$ or $\alpha_{\bf K}$ indicates a
non-analytic behavior of the phonon dispersion, due to a
polynomial decay of the force constants in real space. This
explains why it is impossible for any of the often used
few-nearest-neighbors force constants approaches to properly
describe the HOB phonons near ${\bf K}$ and
${\bm\Gamma}$~\cite{Pavone,Mapelli,Matt,Grun,Sait,Dubay}. The
graphite HOB are almost indistinguishable from those of graphene,
Fig.~\ref{fig1}. The non-analytic behavior at ${\bm \Gamma}$ and
${\bf K}$ is also present in graphite. At ${\bm \Gamma}$ the HOB
is doubly degenerate, consisting of in-plane anti-phase E$_{\rm
2g}$ movements. Near ${\bm \Gamma}$ the two modes split in a upper
longitudinal optical (LO) branch and a lower transverse optical
(TO) branch. $\alpha_{\bm \Gamma}~{\neq\!0}$ only for the LO
branch.
\begin{figure}
\centerline{\includegraphics[width=85mm]{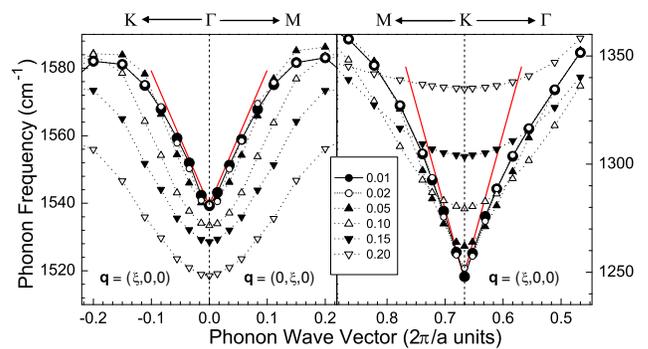}}
\caption{(Color online) Graphene HOB around ${\bm \Gamma}$ and
${\rm \bf K}$ as a function of smearing. Points are calculated
frequencies for $a_{\rm th}$. Lines guides to the eye. The red
straight lines plot Eqs.~\protect\ref{eq10},\protect\ref{eq12}}
\label{fig2}
\end{figure}
\\\indent
Fig.~\ref{fig2} plots the HOB as a function of $\sigma$, to
clarify the nature of the discontinuities. The results for
$\sigma\!=\!0.01~{\rm Ry}$ and $\sigma\!=\!0.02~{\rm Ry}$ are
similar, indicating that, on the scale of the figure, the
$\sigma\!\rightarrow\!0$ limit is reached. Increasing $\sigma$,
the non-analytic behavior is smoothed out. Particularly striking
is the behavior around ${\rm \bf K}$, where for $\sigma >
0.20~{\rm Ry}$ the dispersion is almost flat. Within DFPT, the
smearing $\sigma$ affects virtual transitions between occupied and
empty states, differing in energy by $\lesssim\sigma$. Thus, the
smoothing of Fig.~\ref{fig2} indicates that the HOB
discontinuities are Kohn anomalies~\cite{Kohn59}, since they are
related to an anomalous screening of the electrons around the
Fermi energy.
\\\indent
We compute the EPC matrix elements, to understand why the Kohn
anomalies affect only the HOB and not the others.
For a given phonon mode at the reciprocal-space point
${\rm \bf q}$, we call $\Delta V_{\bf q}$ and $\Delta n_{\bf q}$
the derivative of the Kohn-Sham potential and of the charge density,
with respect to a displacement along the normal coordinate of the phonon.
We define:
\begin{equation} g_{({\rm
\bf k}+{\rm \bf q})i,{\rm \bf k}j} = \langle {\rm \bf k}+{\rm \bf
q},i| \Delta V_{\bf q}{[\Delta n_{\bf q}]} |{\rm \bf k},j\rangle
\sqrt{\hbar/(2M\omega_{\rm \bf q})}, \label{eq1}
\end{equation}
where we consider explicitly the dependence of $\Delta V_{\bf q}$
on $\Delta n_{\bf q}$, as indicated by ${\Delta V_{\bf q}{[\Delta
n_{\bf q}]}}$. $|{{\rm \bf k},i}\rangle$ is the electronic Bloch
eigen-state with wavevector ${\bf k}$, band index $i$, and
eigen-energy $\epsilon_{{\bf k},i}$. $M$ is the atomic mass, and
$g$ is an energy. The dimensionless
EPC~\cite{Gunnarsson} is $\lambda_{\bf q} = {2} \sum_{{\bf k}, i,
j} |g_{({\rm \bf k}+{\rm \bf q})i,{\rm \bf k}j}|^2
\delta(\epsilon_{{\bf k}+{\bf q},i}-\epsilon_{\rm F})
\delta(\epsilon_{{\bf k},j}-\epsilon_{\rm F}) /(\hbar\omega_{\bf
q}N_{\rm F}N_{\rm k})$, where $\sum_{\rm \bf k}$ is a sum on
$N_{\rm k}$ BZ vectors, $N_{\rm F}$ is the density of
states per spin at the Fermi energy $\epsilon_{\rm F}$. In
graphene the Fermi surface is a point, $N_{\rm F}$ is zero and
$\lambda_{\bf q}$ is not well-defined. Thus, we
evaluate~\cite{Gunnarsson}:
\[
\frac{2 \langle g^2_{\bf q}\rangle_{\rm F}}{\hbar\omega_{\bf q}}=
\frac{\lambda_{\bf q}N_{\rm F}}{J_{\bf q}}, \label{eq3}
\]
where $\langle \dots \rangle_{\rm F}$ indicates the average on the
Fermi surface of $|g_{({\rm \bf k}+{\rm \bf q})i,{\rm \bf
k}j}|^2$, and $J_{\bf q}=1/N_k\sum_{i,j,{\bf k}}
\delta(\epsilon_{{\bf k+q},i}-\epsilon_{\rm F})
\delta(\epsilon_{{\bf k},j}-\epsilon_{\rm F})$. In graphene,
$\langle g^2_{\bf K}\rangle_{\rm F} = \sum_{i,j}^\pi |g_{(2{\rm
\bf K})i,{\rm \bf K}j}|^2/4$, and $\langle g^2_{\bm
\Gamma}\rangle_{\rm F} = \sum_{i,j}^\pi |g_{{\rm \bf K}i,{\rm \bf
K}j}|^2/4$, where the sums are performed on the two degenerate
$\pi$ bands at $\epsilon_{\rm F}$. For the HOB, we obtain $\langle
g^2_{\bm \Gamma}\rangle_{\rm F} = 0.0405$~eV$^2$~\cite{jiang}, and $\langle
g^2_{\bf K}\rangle_{\rm F} = 0.0994$~eV$^2$, corresponding to
$2\langle g^2_{\bf q}\rangle_{\rm F}/(\hbar\omega_{\bf q})$ of
$0.41$~eV and $1.23$~eV at ${\bm \Gamma}$ and ${\bf K}$,
respectively. $2 \langle g^2_{\bf K}\rangle_{\rm
F}/(\hbar\omega_{\bf K})$ is much smaller (0.02 eV) for the doubly
degenerate $1200~{\rm cm}^{-1}$ phonon at ${\bf K}$ and zero for
all the other phonons at ${\bf K}$ and at ${\bm \Gamma}$,
consistent with the absence of Kohn anomalies for all these
branches. The EPCs values for the graphene HOB are very large in
absolute terms. They are comparable to the $39~{\rm K}$
superconductor MgB$_2$, for which $\lambda_{\bf q}N_{\rm F}/J_{\bf
q}=$ 1.6 eV~\cite{MgB2us} for each of the two strongly-coupled
E$_{\rm 2g}$ branches at A. On the contrary, in alkaline doped
C$_{60}$, the largest $2 \langle g^2_{\bf q}\rangle_{\rm
F}/(\hbar\omega_{\bf q})$ for individual phonon modes are much
smaller ($0.02$-$0.034$ eV~\cite{Gunnarsson}).
\\\indent
We consider the expression of $\omega_{\bf q}$ according to
perturbation theory ~\cite{DFPT} (within DFT), to understand the
absence of a Kohn anomaly for the TO E$_{\rm 2g}$ phonon at ${\bm
\Gamma}$ and to correlate the constants $\alpha_{\bm \Gamma}$ and
$\alpha_{\bf K}$ with the magnitude of the EPC. $\omega_{\bf
q}=\sqrt{D_{\bf q}/M}$, where $D_{\bf q}$ is the dynamical matrix
projected on the phonon normal coordinate:
\begin{eqnarray}
D_{\bf q} &=& \frac{4}{N_{\rm k}}
\sum_{{\rm \bf k},o,e}
\frac{|
\langle {\rm \bf k}+{\rm \bf q},e|
\Delta V_{\rm \bf q}{[\Delta n_{{\bf q}}]} |{\rm \bf k},o\rangle|^2}
{\epsilon_{{\rm \bf k},o}-\epsilon_{{\rm \bf k}+{\rm \bf q},e}} \nonumber \\
&&-\int
\Delta n^*_{{\rm \bf q}}({\bm r}) K({\bm r},{\bm r'})
\Delta n_{{\rm \bf q}}({\bm r'})~d^3\!r\,d^3\!r' \nonumber \\
&&+\int n({\bm r}) \Delta^2V^b({\bm r})~d^3\!r.
\label{eq4}
\end{eqnarray}
Here a factor 2 accounts for spin degeneracy,
$\sum_{o,e}$ is a sum on occupied and empty states,
$n({\bm r})$ is the charge density,
$K({\bm r},{\bm r'})=\delta^2 E_{Hxc}[n]/
(\delta n({\bm r})\delta n({\bm r'}))$, $E_{Hxc}[n]$ is the Hartree and
exchange-correlation functional,
and $\Delta^2V^b$ is the second derivative of the bare (purely ionic)
potential.  From previous considerations,
\begin{eqnarray}
\alpha_{\bm \Gamma}\! &=& \!\hbar \lim_{{\bf q}\rightarrow 0}
\frac{\omega_{\bf q} - \omega_{\bm \Gamma}}{q}=
\hbar
\lim_{{\bf q}\rightarrow 0}
\frac{D_{{\bf q}} - D_{\bm \Gamma}}{2M\omega_{\bm \Gamma}q},
\label{eq5}\\
\alpha_{\bf K}\! &=&\! \hbar \lim_{{\bf q'}\rightarrow 0}
\frac{\omega_{\bf K+q'} - \omega_{\bf K}}{ q'}=
{\hbar}
\lim_{{\bf q'}\rightarrow 0}
\frac{D_{{\bf K+q'}} - D_{\bf K}}{2M\omega_{\bf K}q'}.
\label{eq6}
\end{eqnarray}
If the dependence of $D_{\bf q}$ on ${\bf q}$ were analytic over
all the BZ, $\alpha_{\bm \Gamma}=\alpha_{\bf K}=0$ (E.g. $D_{{\bf
q}} - D_{\bm \Gamma}={\cal O}(q^2)$). For ${\bf q}$ ${\neq }$
${\bm \Gamma}$ and ${\bf K}$, the energy denominators in the sum
of Eq.~\ref{eq4} are finite and $D_{\bf q}$ is analytic. The
energy denominators go to zero for ${\bf q}={\bm\Gamma}$ (when
${\bf k=\bf K}$ or ${\bf k}$=2${\bf K}$) and for ${\bf q}={\bf K}$
(when ${\bf k=\bf K}$), when $o$ and $e$ correspond to the $\pi$
and $\pi^*$ bands near the Fermi energy. Due to these
singularities, the dynamical matrix is non-analytic for ${\bf
q}={\bf K},~{\bm \Gamma}$, thus $\alpha_{\bm \Gamma}$ and
$\alpha_{\bf K}$ can be different from zero. To compute
$\alpha_{\bf q}$, we can replace in Eqs.~\ref{eq5}-\ref{eq6} the
full dynamical matrix $D_{\bf q}$ with its non-analytic component
$\tilde D_{\bf q}$, which includes only the sum of Eq.~\ref{eq4}
restricted to the $\pi$ electronic bands in an arbitrarily-small
but finite circle of radius $k_{\rm m}$ around the Fermi surface
at the ${\rm \bf K}$ and/or $2{\rm \bf K}$ points~\cite{nota03}.
\\\indent
For ${\bf q}$ near ${\bm \Gamma}$, using the definition of
Eq.~\ref{eq1},
\begin{equation}
\tilde D_{\bf q}=
\frac{8\sqrt{3}M\omega_{\bm \Gamma}}{\hbar}
\!\!\int\limits_{ k'<k_{\rm m}}\!\!\!\!\! d^2\!k'
\frac{|g_{({\bf K+k'+q})\pi^*,({\bf K+k'})\pi}|^2
}{\epsilon_{{\bf K+k'},\pi}-\epsilon_{{\bf K+k'+q},\pi^*}}.
\label{eq7}
\end{equation}
Here we have used the substitution $1/N_{\rm k}\sum_{\bf k} =
\sqrt{3}/2\int d^2\!k'$ (the graphene BZ area is $2/\sqrt{3}$),
${\bf q}$ and ${\bf k}$ points are in units of $2\pi/a$, $a$
being the lattice spacing. ${\bf k'} = {\bf k - K}$. A factor
2 accounts for the contribution of the two equivalent Fermi
points. For small ${\bf q}$ and ${\bf k'}$, the EPC matrix
elements in the numerator in Eq.~\ref{eq7} are:
\begin{equation}
|g_{({\bf K+k'+q})\pi^*,({\bf K+k'})\pi}^{\rm \frac{LO}{TO}}|^2=
\langle g^2_{\bm \Gamma}\rangle_{\rm F}
~[1\pm\cos(\theta+\theta')],
\label{eq8}
\end{equation}
where $\theta$ is the angle between ${\bf k'}$ and ${\bf q}$, and
$\theta'$ is the angle between ${\bf k'}+{\bf q}$ and ${\bf q}$
~\cite{nota04}. The $+$ or $-$ sign refers to the LO and the TO
modes, respectively. In graphene, the electronic bands near the
Fermi level have a conic shape. Therefore, for small ${\bf q}$ and
${\bf k'}$,
\begin{equation}
\epsilon_{{\bf K+k'},\pi}-\epsilon_{{\bf K+k'+q},\pi^*}
= - \beta  k' -\beta |{\bf k'+q}|,
\label{eq9}
\end{equation}
where $\beta =$ 14.1 eV is the slope of the $\pi$ bands within GGA.
Replacing Eq.~\ref{eq7},\ref{eq8}, and \ref{eq9}
in Eq.~\ref{eq5}, one obtains:
\begin{eqnarray}
&&\alpha_{\bm \Gamma}^{\rm  \frac{LO}{TO}} =
\frac{4\sqrt{3}\langle g^2_{\bm \Gamma}\rangle_{\rm F}}
{\beta} \times \nonumber \\
&&\lim_{{\bf q}\rightarrow 0}
\frac{1}{q}\!
\int\limits_{k'<k_{\rm m}}\!\!\!\!\! d^2\! k'
\left[
\frac{1\pm\cos2\theta}
{2k'}
-\frac{1\pm\cos(\theta+\theta')  }
{ k'+|{\bf k'+q}|}
\right] \nonumber \\
&&=
\frac{4\sqrt{3}\langle g^2_{\bm \Gamma}\rangle_{\rm F}}
{\beta} \!
\int\limits_0^\infty\!\! dy \!
\int\limits_0^{2\pi}\!\! d\theta \!
\left[
\frac{1}{2}-
\frac{y\pm y\cos(\theta+\theta')}
{y+\sqrt{1+y^2+2y\cos\theta}}
\right],
\nonumber
\end{eqnarray}
where $y = k/q$, and $\theta'=\arctan [
y\sin\theta/(1+y\cos\theta)]$. The integral is zero for the TO
mode and $\pi^2/4$ for the LO mode. Thus, as expected,
$\alpha^{\rm TO}_{\bm \Gamma}=0$ and
\begin{equation}
\alpha^{\rm LO}_{\bm \Gamma}=
\frac{\langle g^2_{\bm \Gamma}\rangle_{\rm F}}{\beta}
\sqrt{3}\pi^2 = 397~{\rm cm}^{-1}.
\label{eq10}
\end{equation}
For the ${\bf K}$ point, we use an analogous procedure, for
transitions from points in the neighborhood of ${\bf K}$ to points
in the neighborhood of $2{\bf K}$. The EPC matrix elements are:
\begin{equation}
|g_{(2{\bf K+k'+q'})\pi^*,({\bf K+k'})\pi}|^2=
\langle g^2_{\bf K}\rangle_{\rm F}
~(1+\cos\theta''),
\label{eq11}
\end{equation}
with $\theta''$ the angle between ${\bf k'}$ and ${\bf k'}+{\bf
q'}$ ~\cite{nota04}. Eq.~\ref{eq6} becomes:
\begin{equation}
\alpha_{\bf K}=
\frac{2\sqrt{3}\langle g^2_{\bf K}\rangle_{\rm F}}
{\beta}
\lim_{{\bf q'}\rightarrow 0}
\frac{1}{ q'}\!
\int\limits_{k'<k_{\rm m}}\!\!\!\!\! d^2\! k'
\left[
\frac{1}
{k'}
-\frac{1+\cos\theta'' }
{ k'+|{\bf k'+q'}|}
\right].
\nonumber
\end{equation}
The limit of the integral is $\pi^2/2$, and therefore
\begin{equation}
\alpha_{\bf K} =
\frac{\langle g^2_{\bf K}\rangle_{\rm F}}{\beta}
\sqrt{3}\pi^2 = 973~{\rm cm}^{-1}.
\label{eq12}
\end{equation}
The resulting linear dispersions are plotted in Fig.~\ref{fig1}
and Fig.~\ref{fig2}. As expected, the phonon slopes near the
discontinuities are very well reproduced. Finally, we note that,
within a first-neighbors tight-binding (TB) approximation, the
EPCs at ${\bm \Gamma}$ and ${\bf K}$ are not independent, since:
\begin{equation}
\frac{\langle g^2_{\bf K}\rangle_{\rm F}~\omega_{\bf K}}
     {\langle g^2_{\bm \Gamma}\rangle_{\rm F}~\omega_{\bm \Gamma}}
= 2. \label{eq13}
\end{equation}
The validity of this relation is supported by our DFT result, for which
$(\langle g^2_{\bf K}\rangle_{\rm F}~\omega_{\bf K})/
(\langle g^2_{\bm \Gamma}\rangle_{\rm F}~\omega_{\bm \Gamma})=2.02$.
\\\indent
Eqs.~\ref{eq10},\ref{eq12} allow us to directly measure the EPCs at
${\bm \Gamma}$ and ${\bf K}$ from the experimental phonon slopes.
The phonon branches around ${\bm
\Gamma}$ have been determined by several groups with a close
agreement of the measured data~\cite{Pavone}. The fit for
$q\leq~0.15$ of the measurements in Fig.~\ref{fig1}~\cite{Maultzsch04}
with $\hbar\omega_{\bf q}=\hbar\omega_{\bm \Gamma}+\alpha^{\rm
LO}_{\bm \Gamma}q + \gamma q^2$, gives $\alpha^{\rm LO}_{\bm
\Gamma}=340~{\rm cm^{-1}}$, and $\langle g^2_{\bm
\Gamma}\rangle_{\rm F} = 0.035$~eV$^2$~\cite{jiang}. The available data
around ${\bf K}$ are much more scattered. However, from
Eq.~\ref{eq13} we get $\langle g^2_{\bf K}\rangle_{\rm F} =
0.086$~eV$^2$. These values are in excellent agreement with our
calculations and validate our results.
\\\indent
The EPCs near ${\bf K}$ between $\pi^*$ bands~\cite{nota05}
allow the accurate determination of the
intensity and shape of the Raman D peak. This will
be the topic of future publications. Here, we remark that the
A$'_1$ branch has the biggest $\langle g^2_{\bf K}\rangle_{\rm F}$
amongst {\bf K} phonons. Thus, the D peak is due to the highest
optical branch starting from the ${\bf K}$-A$'_1$
mode~\cite{Ferrari00,Maultzsch04,Mapelli,Tuinstra}, not to the branch starting
from the ${\bf K}$-E$'$ mode, as
in~\cite{Matt,Grun,Sait,Pocsik,Thomsen00}. Also, the D peak shifts
linearly with laser excitation energy(${\sim}$50
cm$^{-1}$/eV~\cite{Pocsik}). This is at odds with the flat slope
of the ${\bf K}$-A$'_1$ branch obtained by previous calculations
~\cite{Pavone,Mapelli,Matt,Grun,Sait,Dubay}. But it is consistent
with the ${\bf K}$-A$'_1$ linear-dispersion we get. The D peak
dispersion reflects the slope of the Kohn anomaly at ${\bf K}$ and
provides another independent measure of the EPCs. From
ref.~\cite{Thomsen00}, for $q'\rightarrow 0$, the D peak
dispersion is $\sim\alpha_{\bf K}/\beta$. This gives $\langle
g^2_{\bf K}\rangle_{\rm F} \sim 0.072$~eV$^2$ and, from
Eq.~\ref{eq13}, $\langle g^2_{\bm \Gamma}\rangle_{\rm F} \sim
0.029$~eV$^2$. The EPCs derived in this way are a lower limit
since the experimental D peak dispersion is measured with laser
excitation energies $\ge 1~{\rm eV}$~\cite{Pocsik}. This
corresponds to the phonon slope at $q'\ge 0.035$, which, from
Fig.~\ref{fig2}, underestimates by $30\%$ the slope at $q'=0$.
Taking this into account, the EPCs independently inferred from the
D peak dispersion are as well in excellent agreement with our
calculations.
\\\indent
Due to the reduced dimensionality, we predict even stronger Kohn
anomalies for metallic CNTs, and no anomaly for semiconducting
CNTs. This is the key to differentiate the electrical nature of
CNTs by Raman spectroscopy. A softening of CNT phonons
corresponding to the graphene ${\bm \Gamma}$-E$_{2g}$ mode was
recently reported~\cite{Dubay}. We expect a stronger softening for
the phonons corresponding to the graphene ${\bf K}$-A$'_1$ mode,
since $\langle g^2_{\bf K}\rangle_{\rm F}$$>$$\langle g^2_{\bm
\Gamma}\rangle_{\rm F} $.
\\\indent
In conclusion, we have demonstrated the presence of two remarkable
Kohn anomalies in the phonon dispersions of graphite, revealed by
two kinks for the ${\bm\Gamma}$-E$_{2g}$ and ${\bf K}$-A$'_1$
modes. Even if Kohn anomalies have been observed in many
materials~\cite{DFPT}, graphite is the first real material where a
simple analytic description of the anomaly is possible. Indeed, we
proved, by an exact analytic derivation, that the slope of the
kinks is proportional to the ratio between square EPC matrix
elements and the slope of the $\pi$ bands at the Fermi energy. As
a consequence, we directly derived the EPCs at ${\bm \Gamma}$ and
${\bf K}$ from the experimental phonon dispersions. The values we
obtain are in excellent agreement with calculations. Finally, our
$\langle g^2_{\bm \Gamma}\rangle_{\rm F}$ and $\langle g^2_{\bf
K}\rangle_{\rm F}$ values, with Eqs.~\ref{eq8},\ref{eq11}
and~\cite{nota05}, can be used to determine the mean free path for
electron scattering by optical phonons. This gives the limit of
ballistic transport in CNTs and is of great scientific and
technologic importance~\cite{nanotubes}. This calculation can be
done, within the folding model, using the Fermi golden rule, and
will be reported elsewhere.
\\\indent
We thank C. Brouder, M. Calandra and S. Reich for useful
discussions. Calculations were performed at HPCF (Cambridge) and
IDRIS (Orsay) using PWSCF~\cite{PWSCF}. S.P. was supported by the
EU project FAMOUS and the Marie Curie Fellowship
IHP-HPMT-CT-2000-00209. A.C.F. acknowledges funding from the Royal
Society.


\begin{thebibliography}{99}
\bibitem{nota04}
This angular dependence was derived using a simple tight-binding
model and reproduces very accurately the behavior of our fully
ab-initio calculations.
\end{thebibliography}
\end{document}